\documentclass[prb,article,twocolumn,showpacs]{revtex4}
\usepackage{graphicx}
\usepackage{amsfonts}
\usepackage{amssymb}
\begin{document}

\title{Theory of a peristaltic pump for fermionic quantum fluids}

\begin{abstract}
Motivated by the recent developments in fermionic cold atoms and in nanostructured systems, we propose the model of a peristaltic quantum pump. Differently from the Thouless paradigm, a peristaltic pump is a quantum device that generates a particle flux as the effect of a sliding finite-size microlattice. A one-dimensional tight-binding Hamiltonian model of this quantum machine is formulated and analyzed within a lattice Green's function formalism on the Keldysh contour. The pump observables, as e.g. the pumped particles per cycle, are studied as a function of the pumping frequency, the width of the pumping potential, the particles mean free path and system temperature. The proposed analysis applies to arbitrary peristaltic potentials acting on fermionic quantum fluids confined to one dimension. These confinement conditions can be realized in nanostructured systems or, in a more controllable way, in cold atoms experiments. In view of the validation of the theoretical results, we describe the outcomes of the model considering a fermionic cold atoms system as a paradigmatic example.
\end{abstract}

\author{F. Romeo$^{1}$ and R. Citro$^{1,2}$}
\affiliation{$^{1}$Dipartimento di Fisica "E.R. Caianiello", Universit\`a di Salerno, I-84084 Fisciano (SA), Italy\\
$^{2}$CNR-SPIN Salerno, I-84084 Fisciano (SA), Italy}

\date{\today}
\maketitle

\section{Introduction}
Studying the matter under extreme conditions (of pressure and/or temperature) represents one of the main challenges in modern science and permits, for instance, to clarify the first moments after the Big Bang and the early universe dynamics. A fundamental step towards the full control of temperature dates back to $1908$ when Kamerlingh Onnes first liquefied helium. Since then a sequence of technological and scientific achievements, starting from the discovery of superconductivity \cite{onnes} and superfluidity \cite{Osheroff}, arriving to the Bose-Einstein condensation of alkali atoms \cite{Cornell}, has permitted to test the matter properties very close to the absolute zero, where matter exhibits its quantum nature.\\
Nowadays, more than a century after the beginning of the cryogenic era, ultracold atoms trapped in optical potentials are believed to have the potential to implement universal quantum simulators (UQS) of Hamiltonian models \cite{feynman82,lloyd96,bloch-review2008,weimer2010}. These UQS can be used to study the phase diagrams of several microscopic models by tuning in a very precise way the interaction strength among particles, particles statistics (using bosonic/fermionic atoms), degree of disorder, static or time-dependent potentials, system dimensionality etc.\\
In the last few decades, the extreme flexibility of the cold atoms systems working as UQS has generated an impressive quantity of experimental studies in which the quantum nature of matter plays a crucial role. A recent development concerns quantum transport experiments \cite{brantut2012,krinner2015,husmann2015} in which fermionic reservoirs (with atoms imbalance) are connected by constrictions or short channels which play the role of the quantum point contacts in mesoscopic physics experiments \cite{vanwees88}. Under these non-equilibrium conditions transport properties of the neutral fermionic matter can be studied in a controlled way and concepts, like conductance or mobility, can be easily applied \cite{dattabook}. More interestingly, the concept of quantum pump, first proposed by Thouless \cite{thouless} and intensively studied within a condensed matter context \cite{brouwer98,switkes99,watson}, can be easily implemented in a cold atoms system by means of adiabatic time-modulation of local optical potentials \cite{lohse2016,schweizer2016}.\\
A quantum pump, which can be seen as a particular realization of an adiabatic quantum machine \cite{romeo2009,romeo2010,perroni2014,vonoppen}, is able to transfer particles between distinct reservoirs generating, under appropriate conditions, a quantized particles flux. Since the quantization of the pumped particles number per cycle may be linked to the topological properties of the system, quantum pumps are the object of a renewed theoretical and experimental interest in cold atoms systems \cite{citro2016}.\\
With these motivations, we study here a model of a peristaltic quantum pump, which is a quantum machine generating a particle flux as the effect of a sliding finite-size microlattice. We propose a one-dimensional tight-binding Hamiltonian model which is analyzed using a lattice Green's function formalism on the Keldysh contour \cite{keldysh65}. The pump observables, like e.g. the pumped particles per cycle, are studied as a function of the pumping frequency, the width of the pumping potential, the particles mean free path and the system temperature. The proposed approach allows for the treatment of arbitrary peristaltic potentials and thus is particularly suitable to describe current experiments with fermionic cold atoms \cite{lebrat2017}. Despite the theory is equally applicable to fermionic fluids confined in low-dimensional nanostructures, we discuss the outcomes of the model for a cold atoms system, in which controlled experimental conditions can be realized and spurious effects can be excluded.\\
The paper is organized as follows. In Sec. \ref{sec:HamGF}, we formulate the one-dimensional tight-binding Hamiltonian model of a peristaltic quantum pump, including a detailed discussion on the Keldysh Green's function formalism adopted in the numerics. Numerical results and their relevance for experiments are analyzed in Sec. \ref{sec:Results}. Conclusions are reported in Sec. \ref{sec:Conclusions}. Details on the theoretical treatment are given in Appendices A, B, C, D, E.

\section{Hamiltonian model and Green's functions theory}
\label{sec:HamGF}
\begin{figure}
\includegraphics[clip,scale=0.45]{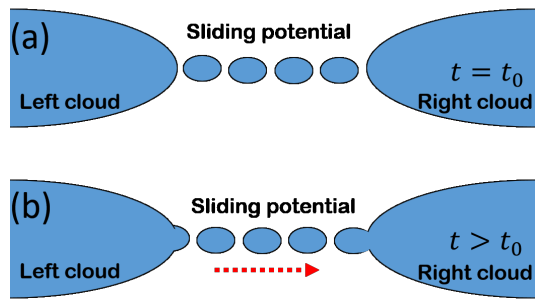}
\caption{(a) Schematic representation of the peristaltic quantum pump realized with an ultracold gas of fermionic atoms. The application of the static potential splits the atomic cloud into two separate parts, namely the left and the right cloud. The central region is characterized by a depleted atomic density on which the peristaltic potential (sliding potential) is imprinted. The sliding part of the potential induces a momentum transfer to the atomic gas responsible for a net particle current as shown in the lower panel (b).}
\label{fig:fig1}
\end{figure}

\begin{figure}
\includegraphics[clip,scale=0.45]{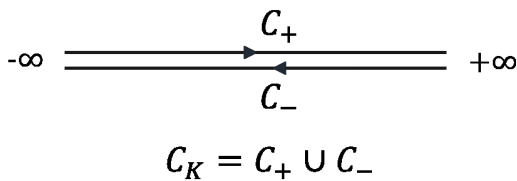}
\caption{The Keldysh contour $\mathcal{C}_{K}=C_{+}\cup C_{-}$ made of two counter-propagating copies of the time axis, with $C_{+}$ ($C_{-}$) parallel (antiparallel) to the ordinary time axis.}
\label{fig:fig2}
\end{figure}

We consider a fermionic (spinless) fluid confined in one dimension which can be realized with high degree of control using fermionic ultracold gases. In the absence of external potentials, the tight-binding Hamiltonian
\begin{equation}
H_0=\sum_n \epsilon c^\dagger_n c_n-J\sum_n (c^\dagger_{n+1}c_n+h.c.)
\end{equation}
describes delocalized particles on the lattice characterized by on-site energy $\epsilon$, hopping energy $J>0$ and lattice constant $a$. The fermionic nature of the particles is encoded by the creation and annihilation operators, $c^{\dagger}_{n}$ and $c_{n}$ respectively, whose anticommutation relations are $\{c_n,c^{\dagger}_m\}=\delta_{nm}$ and $\{c_n,c_m\}=\{c^{\dagger}_n,c^{\dagger}_m\}=0$. The one dimensional fluid can be perturbed introducing a localized time-dependent potential on a finite-length region $\mathcal{A}$, represented by a finite subset of the lattice sites. In the presence of the perturbation, the system Hamiltonian becomes $H=H_0+H_V$, where
\begin{equation}
H_V=\sum_{j\in \mathcal{A}} V_j(t) n_j
\end{equation}
describes the interaction between the external potential $V_j(t)=V^0_j+V_j \cos(\omega t+ \varphi_j)$ and the particle number operator $n_j=c^{\dagger}_{j}c_{j}$. The static component of the potential $V^0_j$ strongly depletes the fluid at $j \in \mathcal{A}$; this region of depleted density splits the fermionic gas in two separate clouds, located on the left and on the right of the static barrier. Particles belonging to the left cloud can be transferred to the right cloud (or viceversa) using the peristaltic potential $V_j \cos(\omega t+ \varphi_j)$, with $\varphi_j=Kaj$ (see Fig. \ref{fig:fig1}). The particle current generated by the pump presents an ac contribution (whose average over a pumping period is zero) and a dc component. The latter contribution is related to the transfer efficiency of the pump which is measured as the number of atoms pumped within a pumping period $2\pi/\omega$. Hereafter, we formulate a Keldysh Green's functions theory and discuss the efficiency of a peristaltic pump, which is a measurable quantity in the experiments with cold atoms.\\
The bond current flowing at site $l$ is obtained from Heisenberg equation of motion for the number operator $n_l=c^\dagger_l c_l$ and taking its
quantum-statistical average:
\begin{equation}
\partial_t \langle n_l \rangle= \frac{-i}{\hbar}\langle[n_l, H]\rangle=-\frac{2 J}{\hbar}\sum_{r=\pm 1} Re \lbrack G^{<}_{l,l+r}(t,t)  \rbrack,
\end{equation}
where we have introduced the lesser Green's function:
\begin{equation}
\label{eq:lesser}
G^{<}_{nm}(t,t')=i \langle c^\dagger_m (t') c_{n}(t)\rangle.
\end{equation}
The lesser Green's function $G^{<}$ together with the advanced/retarded Green's functions, $G^{A/R}$, provide a suitable set of correlation functions to characterize non-equilibrium phenomena. In the following, we briefly recall the main steps required to calculate the non-equilibrium properties.\\
The mathematical approach used in developing a non-equilibrium theory requires the deformation of the time axis to the so-called Keldysh contour $\mathcal{C}_K$, which is a closed time path over which time ordering $T_{C}[...]$ and time variables $\tau$ are defined. The contour-ordered Green's function $G_{nm}(\tau,\tau')\equiv -i \langle T_{C}[c_{n}(\tau)c_{m}^{\dagger}(\tau')]\rangle$ can be defined, with the time arguments $\tau,\tau' \in \mathcal{C}_{K}$ defined in the Keldysh contour. The contour-ordered Green's function admits a perturbation
expansion which is mathematically equivalent to the Dyson equation of the equilibrium case. The Dyson equation on the Keldysh contour takes the familiar form:
\begin{eqnarray}
\label{eq:dysonK}
G=g+gVG+g\Sigma G,
\end{eqnarray}
where $g$ is the unperturbed Green's function, $V$ represents a single-particle potential and $\Sigma$ is the single-particle irreducible self-energy. In the Eq. (\ref{eq:dysonK}) the shortened notation $g\Sigma G$ stands for:
\begin{eqnarray}
&&[g\Sigma G]_{nm}(\tau,\tau')=\\\nonumber
&=&\sum_{ls}\int_{\mathcal{C}_{K}}d\tau_a d\tau_b \ g_{nl}(\tau,\tau_a)\Sigma_{ls}(\tau_a,\tau_b)G_{sm}(\tau_b,\tau'),
\end{eqnarray}
where $n,m$ are lattice indices. For noninteracting particles, which is the case of interest for the subsequent discussion, $\Sigma_{ls}(\tau_a,\tau_b)\rightarrow 0$ and the Dyson equation takes the form:
\begin{eqnarray}
G_{nm}(\tau,\tau')&=&g_{nm}(\tau,\tau')+\\\nonumber
&+&\sum_{l}\int_{\mathcal{C}_{K}}d\tau_a \ g_{nl}(\tau,\tau_a)V_{l}(\tau_a)G_{lm}(\tau_a,\tau'),
\end{eqnarray}
with $V_{l}(\tau_a)=V^0_l+V_l \cos(\omega \tau_{a}+ \varphi_l)$ the single particle potential. The contour Dyson equation needs to be projected on the ordinary time axis. This projection is not trivial since the Keldysh contour $\mathcal{C}_{K}=C_{+}\cup C_{-}$ (see Fig. \ref{fig:fig2}) is made of two counter-propagating copies of the time axis, i.e. $C_{+}$ and $C_{-}$, and thus a two-time correlation $G(\tau,\tau')$ defined on the contour generates four real-time correlations when projected on the ordinary time axis. Indeed, since $\tau, \tau'$ can belong either to $C_{+}$ or $C_{-}$, four different choices are possible.
In this way Keldysh theory presents a matrix structure with respect to the time branches:
\begin{eqnarray}
G(\tau,\tau')\Leftrightarrow\left(
                               \begin{array}{cc}
                                 G_{++}(t,t') & G_{+-}(t,t') \\
                                 G_{-+}(t,t') & G_{--}(t,t') \\
                               \end{array}
                             \right),
\end{eqnarray}
where, for instance, $G_{+-}(t,t')$ represents the projection of the contour correlation $G(\tau,\tau')$ on the ordinary time axis by fixing the contour variables as $\tau \in C_{+}$ and $\tau' \in C_{-}$. Only three of the four correlations are independent and we have the freedom of formulating the entire theory in terms of the advanced correlation $G^{A}(t,t')$, the retarded correlation $G^{R}(t,t')$ and the lesser Green's function $G^{<}(t,t')$. In this way, the projection on the real time axis of the contour Dyson equation is made by substituting the generic two-times correlation $B(\tau,\tau')$ on the contour by the real-time quantity:
\begin{eqnarray}
\label{eq:mat1}
                              \left(
                               \begin{array}{cc}
                                 B^{R}(t,t') & B^{<}(t,t') \\
                                 0 & B^{A}(t,t') \\
                               \end{array}
                             \right),
\end{eqnarray}
while the integration on the contour variable $\int_{\mathcal{C}_{K}} d\tau$ is substituted by ordinary integration on time $\int dt$. Quantities which depend on one time variable, e.g. $V(\tau)$, are substituted by:
\begin{eqnarray}
\label{eq:mat2}
                              \left(
                               \begin{array}{cc}
                                 V(t) & 0 \\
                                 0 & V(t) \\
                               \end{array}
                             \right),
\end{eqnarray}
since they are diagonal in the time branch indices ($+,-$). Once the above substitutions have been made, the contour Dyson equation generates three distinct equations for the retarded, advanced and the lesser part of the Green's function. Within a matrix representation in the lattice indices, the advanced and retarded part of the Green's function $\hat{G}^{R,A}$ follow from the usual form of the Dyson equation:
\begin{eqnarray}
\label{eq:gretav}
\hat{G}^{R,A}(t,t')&=&\hat{g}^{R,A}(t,t')+\\\nonumber
&+&\int dt_a\hat{g}^{R,A}(t,t_a)\hat{V}(t_a)\hat{G}^{R,A}(t_a,t'),
\end{eqnarray}
where we introduced the matrix $\hat{V}(t_a)$ of elements $V_{n}(t_a)\delta_{nm}$ representing the potential, while $\hat{g}^{R,A}(t,t')$ represent the unperturbed correlations computed according to Appendix \ref{app:gfDos}. The lesser Green's function $\hat{G}^{<}$ obeys the integral equation:
\begin{eqnarray}
\label{eq:glimpl}
\hat{G}^{<}(t,t')&=&\hat{g}^{<}(t,t')+\\\nonumber
&+&\int dt_a\hat{g}^{R}(t,t_a)\hat{V}(t_a)\hat{G}^{<}(t_a,t')+\\\nonumber
&+&\int dt_a\hat{g}^{<}(t,t_a)\hat{V}(t_a)\hat{G}^{A}(t_a,t'),
\end{eqnarray}
with $\hat{g}^{<}(t,t')$ the unperturbed lesser Green's function (see Appendix \ref{app:lesser} for details).
Equation (\ref{eq:glimpl}) admits the following formal solution for $G^{<}$
\begin{eqnarray}
\label{eq:glexpl}
\hat{G}^{<}(t,t')&=&\hat{g}^{<}(t,t')+\\\nonumber
&+&\int dt_a\hat{g}^{<}(t,t_a)\hat{V}(t_a)\hat{G}^{A}(t_a,t')+\\\nonumber
&+&\int dt_a\hat{G}^{R}(t,t_a)\hat{V}(t_a)\hat{g}^{<}(t_a,t')+\\\nonumber
&+&\int dt_a dt_b\hat{G}^{R}(t,t_a)\hat{\Pi}^{<}(t_a,t_b)\hat{G}^{A}(t_b,t'),
\end{eqnarray}
coming from infinite order iteration of Eq. (\ref{eq:glimpl}) with the definition of the lesser self-energy $\hat{\Pi}^{<}(t_a,t_b)=\hat{V}(t_a) \hat{g}^{<}(t_a,t_b) \hat{V}(t_b)$. Equations (\ref{eq:gretav}) and (\ref{eq:glexpl}) complete the projection of the contour Dyson equation on the real-time axis and fully define the non-equilibrium properties of the system. Moreover the matrix formalism defined in Equations (\ref{eq:mat1}) and (\ref{eq:mat2}) automatically implements the Langreth rules \cite{langreth91} for the so-called \textit{analytic continuation} (see Appendix \ref{app:langreth}) from the Keldysh contour to the real-time axis.\\

The calculation of the dc current generated by the peristaltic pump requires at least the second order approximation of the lesser Green's function with respect to the pumping potential. Thus, in the following we explain the approximation scheme adopted in the computation. First of all, we rewrite the single-particle potential in the form $\hat{V}(\tau)=\hat{V}_0+\hat{V}_P(\tau)$, where $\hat{V}_0$ represents the static part of the potential with component $[\hat{V}_0]_{ln}=V^{0}_{l}\delta_{ln}$ in the lattice sites representation, while $\hat{V}_P(\tau)$ is the time-dependent part of the pumping potential whose components are $[\hat{V}_P(\tau)]_{ln}=V_{l}\cos(\omega \tau+\varphi_l)\delta_{ln}$. Under the assumption of a localized perturbation of moderate modulation amplitude (i.e., $V^{0}_{l}>V_{l}$, $\forall l \in \mathcal{A}$), we approximate the retarded/advanced Green's function by the exact solution of the Dyson equation $\mathcal{G}^{R/A}=g^{R/A}+g^{R/A}V_0\mathcal{G}^{R/A}$, in which the time-dependent part of the potential $\hat{V}_P$ has been neglected. The Green functions $\mathcal{G}^{R/A}$ describe the quantum correlations of two atomic clouds separated by a static potential barrier (reservoirs) under the assumption (to be validated at the end of the computation) that the time-dependent part of the potential does not perturb the bulk of the atomic reservoirs. Substituting $\mathcal{G}^{R/A}$ in Eq. (\ref{eq:glexpl}) we obtain a second order approximation of $G^{<}$ with respect to the pumping potential. Three types of terms are generated: (i) terms which depend only on the static part of the potential; (ii) linear terms in the pumping potential; (iii) quadratic terms in the pumping potential. Terms in (i) only provide a small correction to the unperturbed density of states of the system and do not induce particle current once the renormalized chemical potential is considered. Terms in (ii), essentially generated by the convolution of the forcing term with the Lindhard response function \cite{flensbergbook}, are responsible for the ac linear response to the pumping potential and provide a time-dependent current with vanishing average over a pumping cycle. Terms in (iii) characterize the lowest-order non-linear response of the system to the pumping signal. The non-linear response, which is described by
\begin{equation}
\label{eq:keldysh}
\delta \hat{G}^{<}(t,t)=\int dt_1 dt_2\hat{\mathcal{G}}^R(t,t_1)\hat{\Pi}_{P}^<(t_1,t_2)\hat{\mathcal{G}}^A(t_2,t),
\end{equation}
with $\hat{\Pi}_{P}^<(t,t')=\hat{V}_P(t) \hat{g}^{<}(t,t') \hat{V}_P(t')$, generates the particle current:
\begin{equation}
\mathcal{J}_n(t)=-\frac{2 J}{\hbar}\sum_{r=\pm 1} Re \lbrack \delta G^{<}_{n,n+r}(t,t)\rbrack,
\end{equation}
containing a zero-average second-harmonics term and a dc term $\mathcal{J}^{dc}_n$,
\begin{eqnarray}
\mathcal{J}^{dc}_n=\frac{\omega}{2\pi}\int_{0}^{2\pi/\omega}\mathcal{J}_n(t) dt.
\end{eqnarray}
The dc current $\mathcal{J}_L$ ($\mathcal{J}_R$) pumped in the left (right) side of the one dimensional channel is given by $\mathcal{J}_L=\sum_{n \in L}\mathcal{J}^{dc}_n$ ($\mathcal{J}_R=\sum_{n \in R}\mathcal{J}^{dc}_n$) and the current conservation implies $\mathcal{J}_R=-\mathcal{J}_L$. The efficiency of the peristaltic pump, which is the object of our analysis, is defined as the number of atoms pumped during a pumping cycle: $\mathcal{N}_{L/R}=2\pi \mathcal{J}_{L/R}/\omega$. The above equations make a link between the physical observable $\mathcal{N}_{L/R}$, defining the efficiency of the pump, and the correlation function $\delta \hat{G}^{<}(t,t)$. \\
The quantity $\delta \hat{G}^{<}(t,t)$ can be expressed via the Fourier transformation (see Appendix \ref{app:fourier})
\begin{equation}
\label{eq:keldysh}
\delta \hat{G}^{<}(t,t)=\int \frac{dE_1 dE_2}{(2\pi \hbar)^2}\delta \hat{G}^{<}(E_1,E_2)e^{-i \frac{(E_1-E_2)}{\hbar}t},
\end{equation}
in which the Fourier transform $\hat{G}^{<}(E_1,E_2)$ obeys the equation:
\begin{eqnarray}
\label{eq:g_lesser_en}
&&\delta \hat{G}^{<}(E_1,E_2)=\\\nonumber
&&=\int \frac{dE_a dE_b}{(2\pi \hbar)^2}\hat{\mathcal{G}}^R(E_1,E_a)\hat{\Pi}_{P}^<(E_a,E_b)\hat{\mathcal{G}}^A(E_b,E_2).
\end{eqnarray}
The Green's functions $\hat{\mathcal{G}}^{R/A}$ are translational invariant in time, i.e. $\mathcal{G}_{nm}^{R,A}(E_1,E_2)=2\pi \hbar \delta(E_1-E_2) \mathbb{G}^{R,A}_{nm}(E_1)$, since they are exact solutions of the Dyson equation in which the time-dependent part of the potential has been neglected. The information on the time dependent part of the potential is instead encoded in the lesser self-energy which takes the following form:
\begin{eqnarray}
\label{eq:pi}
&&[\hat{\Pi}_{P}^<(E_a,E_b)]_{nm}=\\\nonumber
&&=(2\pi\hbar)^2\sum_{k;\eta,\eta'=\pm}\frac{i\mathcal{A}^{(\eta)}_n \mathcal{A}^{(\eta')}_m}{N}f(E_k)e^{ika(n-m)} \times\\\nonumber
&&\delta(E_a+\eta \hbar \omega-E_k)\delta(E_a-E_b+\hbar \omega(\eta+\eta')),
\end{eqnarray}
where the complex pumping amplitudes $\mathcal{A}^{(\eta)}_n=(V_{n}/2) e^{i\eta Kan}$ have been introduced. Here the quantity $K$ represents the momentum transferred by the pumping potential to the atomic cloud and originates from the phases $\varphi_n=K a n$ of the peristaltic potential, while the indices $\eta,\eta'=\pm$ are related to atom-photon scattering processes involving the emission/absorption of photons with energy $\hbar\omega$. \\
Using (\ref{eq:pi}) in (\ref{eq:g_lesser_en}) and working in time domain one obtains an expression for the correction to the lesser Green's function of the form:
\begin{eqnarray}
\delta \hat{G}^{<}(t,t)=\sum_{\eta,\eta'=\pm}e^{i(\eta+\eta')\omega t}\hat{\mathbb{C}}(\eta,\eta'),
\end{eqnarray}
whose time average $\overline{\delta \hat{G}^{<}}$ is obtained by retaining the terms with $\eta'=-\eta$ in the summation; the auxiliary function $\hat{\mathbb{C}}(\eta,-\eta)$ is given by the expression:
\begin{equation}
\hat{\mathbb{C}}(\eta,-\eta)=\sum_{k} \hat{\mathbb{G}}^R(E_k-\eta \hbar \omega)\hat{\mathcal{M}}^{(\eta)}(k,K)\hat{\mathbb{G}}^A(E_k-\eta \hbar \omega),
\end{equation}
in which the notation $[\hat{\mathcal{M}}^{(\eta)}(k,K)]_{nm}=i\frac{V_n V_m}{4 N}f(E_k)e^{i(k+\eta K)(n-m)a}$ has been introduced. Finally, the space distribution of the atom current can be derived according to the following relation:
\begin{equation}
\mathcal{J}^{dc}_n=-\Omega_0 \sum_{r=\pm 1} Re \Big\{ \Big(\overline{\delta \hat{G}^{<}}\Big)_{n,n+r}\Big\},
\end{equation}
with $\Omega_0=2J/\hbar$.

\section{Numerical results}
\label{sec:Results}
Up to now we have considered a quite general single particle potential of the form $\hat{V}(t)=\hat{V}_0+\hat{V}_P(t)$, including a static part, $\hat{V}_0$, with components $[\hat{V}_0]_{ln}=V^{0}_{l}\delta_{ln}$, and a time-dependent part, $\hat{V}_P(t)$, whose components are $[\hat{V}_P(t)]_{ln}=V_{l}\cos(\omega t+\varphi_l)\delta_{ln}$. We now define a potential profile, which is easily realized with optical methods in cold atoms systems, making a specific choice for $V_l$ and $V_l^{0}$. In particular, the static part of the potential, which is responsible for the splitting of the atomic cloud into a left and a right cloud, is modeled as a rectangular barrier defined as $V^{0}_{n}=V$ for $n \in [n_0-W,n_0+W]\equiv \mathcal{A}$ and $V^{0}_{n}=0$, elsewhere. On the other hand, we introduce a parabolic envelope of the time-dependent part which is different from zero only for $n \in \mathcal{A}$ and its functional form is defined by the relation $V_n=U_{P}\left[1-(\frac{n-n_0}{W})^2\right]$. Consequently, the resulting potential is centered at the lattice position $n_{0}$, while the width $2W$ defines the distance between the two atomic clouds. In order to obtain a system made of two weakly coupled clouds (atomic reservoirs), we set $V>E_F$ and $2W>\lambda_F$, with $E_F$ and $\lambda_F$ the Fermi energy and the Fermi wavelength, respectively. Under these conditions the static part of the potential strongly depletes the particles density in $\mathcal{A}$, which is partially restored by the action of the time-dependent part of the potential.
The time-dependent deformation of the potential profile generates a particle current in  $\mathcal{A}$ exploiting the same working principle of a peristaltic pump for ordinary fluids and thus, in the following, we designate this kind of pumps as \textit{peristaltic quantum pumps}. Similarly to the approach proposed by P. W. Brouwer \cite{brouwer98}, the perturbative treatment of the pumping potential developed in Sec. \ref{sec:HamGF} requires the condition $U_P<V$.\\
As will be clear in the following, the physics of the system under investigation is strongly affected by the presence of quantum states localized inside the region $\mathcal{A}$, where the interplay of the static and time-dependent part of the potential defines a sliding microlattice. Such states, describing particles trapped inside local minima of the microlattice, are coupled to the atomic reservoirs and can be occupied by atoms loaded during the peristaltic motion of the pump.
\begin{figure}[h!]
\includegraphics[clip,scale=0.75]{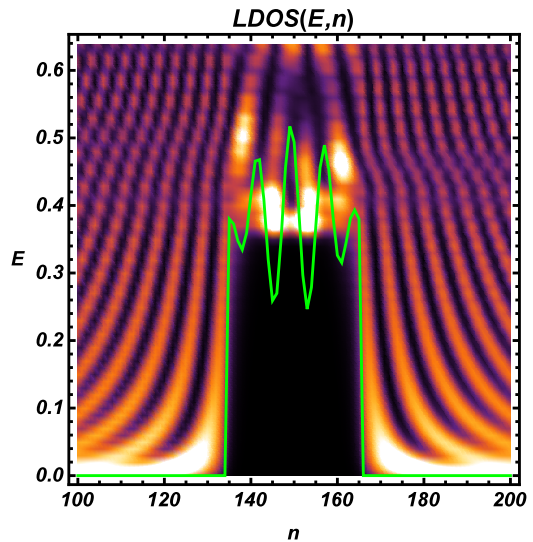}
\caption{Local density of states $LDOS(E,n)$ of a one dimensional lattice with $N=300$ lattice sites with hopping integral $J$ and onsite energy $\epsilon=2J$. The energy $E$ is expressed in units of the hopping $J$, while $n$ represents the position of the $n$-th lattice site. Full line (green in color version) represents the single particle potential with $V=0.38J$, $U_P=0.14J$, $Ka=0.8$, $W=15$, $n_0=150$ and $t=0$. Darker (brighter) regions represent lower (higher) values of $LDOS(E,n)$. High values of LDOS, corresponding to localized quantum states, are evident inside the barrier region (i.e. $n \in [135,165]\equiv \mathcal{A}$) for energy $E \approx 0.4 J$.}
\label{fig:fig3}
\end{figure}
In particular, the alinement among the energy levels of the mentioned quantum states and the system Fermi energy strongly affects the pumping efficiency. In order to define the fingerprint of these states, in Fig. \ref{fig:fig3} we study the local density of states $LDOS(E,n)$ of a one dimensional lattice with $N=300$ lattice sites under the action of the (time-independent) single-particle potential $\hat{V}(t=0)$, representing the full potential $\hat{V}(t)$ frozen at the initial time $t=0$. For $n \in \mathcal{A}$ (i.e. the region where the potential is different from zero) the potential $\hat{V}(t=0)$ is parametrized as
\begin{eqnarray}
&&[\hat{V}(t=0)]_{ln}=\\\nonumber
&=&\Big\{V+U_{P}\left[1-\Big(\frac{n-n_0}{W}\Big)^2\right]\cos(Kan)\Big\}\delta_{ln},
\end{eqnarray}
with $V=0.38J$, $U_P=0.14J$, $Ka=0.8$, $W=15$ and $n_0=150$. The effect of the potential is twofold. Far from the influence region of the potential the interference of back scattered and incident particles produces modulation of the local density of states whose spatial period depends on the energy $E$ of the scattered atoms. Inside the scattering region $\mathcal{A}$ and for $E<0.38J$ the local density of states is strongly suppressed due to the barrier effect of the potential, while for $E \in [0.38 J,0.55 J]$ localized quantum states (describing atoms trapped inside the minima of the microlattice) are responsible for an increased density of states. These states are involved in resonant tunneling events which play a crucial role for the pump operation (see Appendix \ref{app:scattering microlattice}). The subsequent analysis shows the relevance of these states under dynamic conditions, i.e. when the peristaltic pump is working. For all the simulations results, current conservation law $\mathcal{J}_L+\mathcal{J}_R=0$ is always respected within the numerical error determined by the machine precision (typically $\sim 10^{-15}$ using dimensionless units).\\
\begin{figure*}[t]
\includegraphics[clip,scale=0.75]{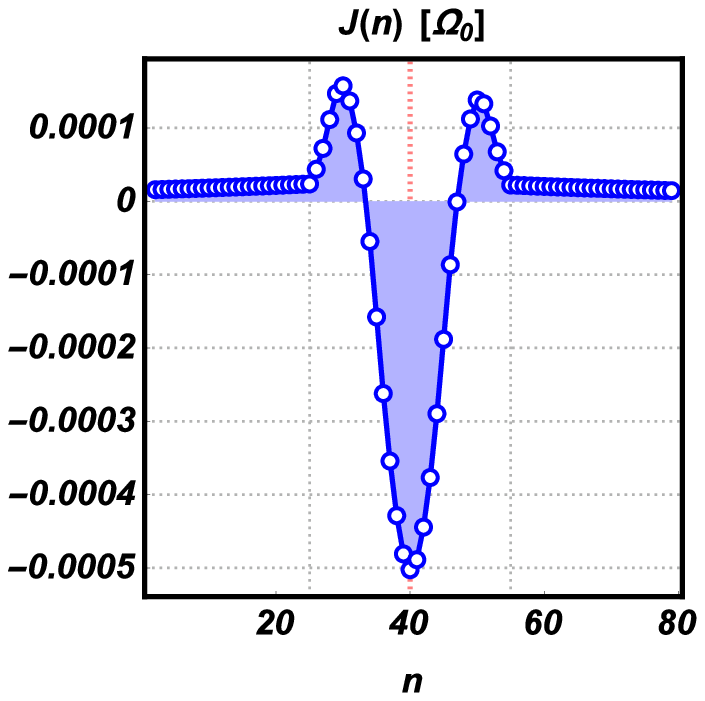}\includegraphics[clip,scale=0.75]{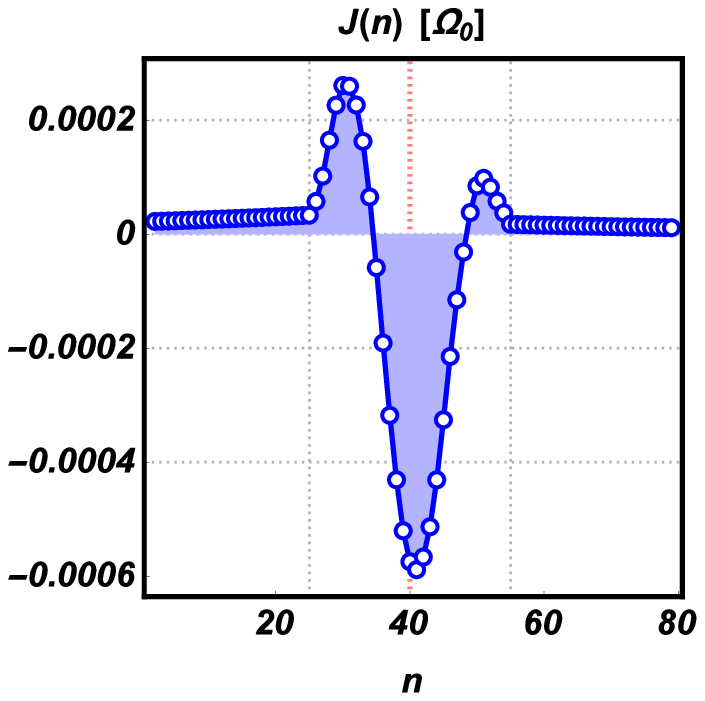}\includegraphics[clip,scale=0.78]{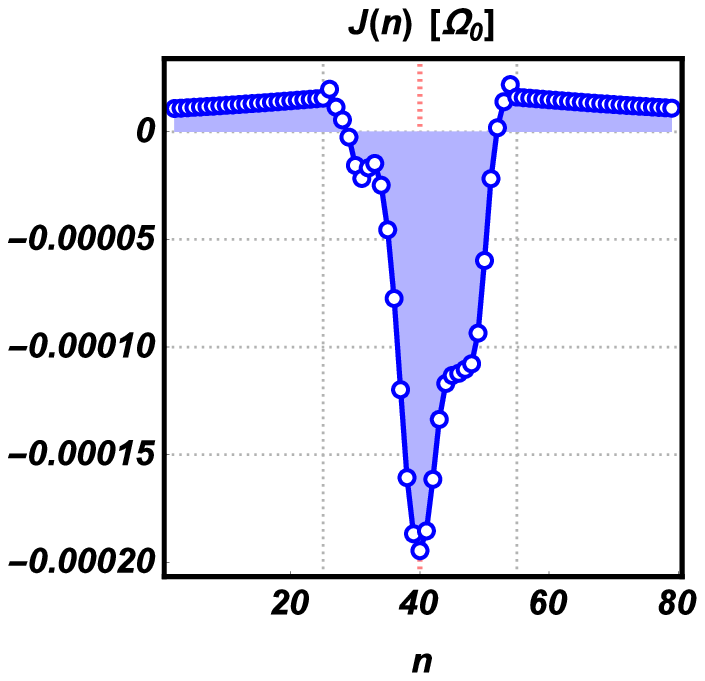}
\caption{Dimensionless bond current $\mathrm{J}(n)\equiv\mathcal{J}_n^{dc}/\Omega_0$ as a function of the lattice position $n$ for a one dimensional lattice with $N=80$ sites. The simulation parameters in all panels are fixed as: $\epsilon=2J$, $V=0.385J$, $U_P=0.14J$, $W=15$, $n_0=40$, $E_F \approx 0.382 J$, $k_B T=0.1 \cdot E_F$, $\Gamma=0.01J$. The remaining parameters are fixed as $Ka=0.8$ and $\hbar \omega/J=5\cdot 10^{-4}$ for the left panel and $Ka=0.8$ and $\hbar \omega/J=5\cdot 10^{-3}$ for the middle panel. Right panel has been computed using $Ka=1$ and $\hbar \omega/J=5\cdot 10^{-3}$. Vertical dashed lines define the boundaries of the microlattice region. The bond current is mainly generated within the microlattice region and strongly depends on the pumping parameters $Ka$ and $\omega$.}
\label{fig:fig4}
\end{figure*}
We have simulated a one dimensional system consisting of $N=80$ lattice sites, which represents the minimal setting to minimize finite size effects. As will be clear in the following discussion, the simulated region can also be imagined as the active region (i.e. the region where the current is generated) of a bigger system. The on site energy has been fixed to $\epsilon=2J$ so that the resulting energy band spans the energy range $E \in [0,4J]$. The Fermi energy has been fixed to $E_F \approx 0.382 J$ to obtain a Fermi wavelength of ten lattice sites ($\lambda_F \approx 10 a$), which corresponds to the microlattice spatial period when $Ka \approx 0.8-1.0$. The width of the barrier $2W$, splitting the atomic cloud into two separate subsystems, is normally fixed to $30$ lattice sites, which is appropriate to deplete the local density of states inside the scattering region. Since we are interested in the description of a quantum machine working close to the adiabatic regime, we consider pumping frequencies in the range $\hbar \omega/J=10^{-3}-10^{-2}$. The study of the $\mathcal{J}_{L/R}$ versus $\omega$ curves (not reported here) clearly evidences the role of finite frequency effects which manifest itself as deviations from the linear behavior expected in the adiabatic regime. Such effects strongly depend on the potential profile parameters (e.g. $Ka$, $V$, etc.) and on the pump working point. The system temperature has been fixed to $k_B T=0.1 \cdot E_F$, which is appropriate to describe current experiments with fermionic cold atoms. A finite particles relaxation time $\tau$ is also included in the model by introducing the phenomenological parameter $\Gamma=\hbar/(2 \tau)$ (see Appendix \ref{app:gfDos}), while the mean free path $\ell$ associated to the relaxation time $\tau$ can be estimated using the relation $\ell \approx v_F \tau$, being $v_F$ the Fermi velocity of the system. Under the assumption of a parabolic dispersion relation and introducing the quantity $\zeta=\lambda_F/(2W)<1$ as the dimensionless ratio between the Fermi wavelength $\lambda_F$ and the barrier width $2W$, we derive the useful relation $\ell/(2W)=\zeta E_F/(2\pi \Gamma)$. Setting the typical simulation values $\zeta=1/3$, $E_F=0.382J$ and $\Gamma=0.01J$, we get $\ell/(2W) \approx 2$, which ensures ballistic transport through the scattering region. From the experimental viewpoint, a finite mean free path for a fermionic atom can originate from the non-elementary nature of such particles. Indeed, under specific dynamical regimes, the coupling between internal degrees of freedom with the center of mass motion of the atom may well mimic \textit{scattering} events responsible for a finite mean free path.\\
First of all we observe that the bond current, which is the result of the peristaltic motion of the pump, is mainly generated within the microlattice region and strongly depends on the pumping parameters $K$ and $\omega$. This behavior, which validates the approximations used in the computation of the reservoirs Green functions $\mathcal{G}^{R/A}$, is evident in Figure \ref{fig:fig4} where we report the dimensionless bond current $\mathrm{J}(n)\equiv\mathcal{J}_n^{dc}/\Omega_0$ along the one dimensional system. \\
The left and middle panel of Fig. \ref{fig:fig4} are obtained by fixing the same value of the microlattice spacing ($Ka=0.8$), while increasing the pumping frequency of one order of magnitude going from the left to the middle panel.
The comparison of those figures shows that asymmetries in the spatial distribution of the bond current are expected when the pumping frequency is increased and finite frequency effects starts to affect the pump operation. The analysis of the middle and right panel of Fig. \ref{fig:fig4}, which are obtained by fixing the same value of the pumping frequency ($\hbar \omega/J=5\cdot 10^{-3}$) while taking $Ka=0.8$ ($Ka=1$) for the middle (right) panel, evidences that asymmetries in the spatial distribution of the bond current are enhanced when the microlattice spacing is decreased (i.e. when the pump parameter $Ka$ is increased). This behavior reflects the departure from the adiabatic regime and does not depend on the specific choice of simulation parameters.
The distribution of the bond current produces observable effects on the efficiency of the peristaltic pump. The pump efficiency depends in a complicated way on all the relevant parameters characterizing the peristaltic potential, however, from the experimental viewpoint, gate potentials are applied to the system in order to modify the pump working point and thus the transfer efficiency of the pump. Gate potentials in cold atoms systems, like the electrostatic gates in nanostructures, produce a change in the density of states of the channel and thus lead to a chemical potential adjustment.
\begin{figure}[h]
\includegraphics[clip,scale=0.85]{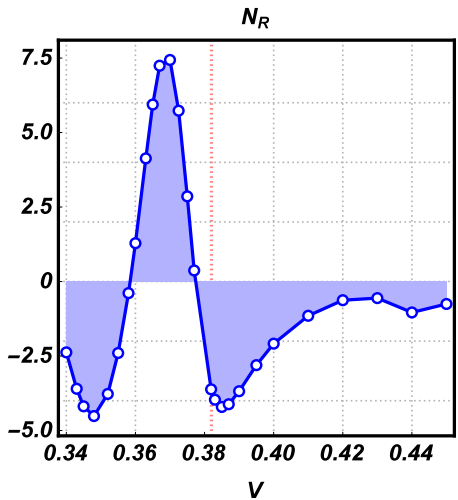}
\caption{Number of atoms $\mathcal{N}_R$ pumped in a pumping cycle as a function of the static barrier $V$ (in units of $J$) obtained by fixing the model parameters as follows: $\epsilon=2J$, $\hbar \omega/J=5\cdot 10^{-3}$, $U_P=0.14J$, $Ka=0.8$, $W=15$, $n_0=40$, $N=80$, $E_F \approx 0.382 J$, $k_B T=0.1 \cdot E_F$, $\Gamma=0.01J$. Positive (negative) values of $\mathcal{N}_R$ indicate that the pump works adding (removing) $|\mathcal{N}_R|$ atoms per cycle to the right atomic cloud.}
\label{fig:fig5}
\end{figure}
In our model, the effect of a moderate chemical potential adjustment can be emulated by changing the static barrier $V$, while keeping fixed the chemical potential. Thus, in Fig. \ref{fig:fig5} we study the number of atoms $\mathcal{N}_R$ pumped in a pumping cycle as a function of the static barrier $V$.
Within the relevant $V$ range, a strong modulation of the pump efficiency as a function of $V$ is observed accompanied by changing of sign of the $\mathcal{N}_R$ \textit{vs} $V$ curve. From the physical viewpoint, positive (negative) values of $\mathcal{N}_R$ indicate that the pump works adding (removing) $|\mathcal{N}_R|$ atoms per cycle to the right atomic cloud. The transfer efficiency of the system is maximized for $V \approx 0.37J$, which is very close to the system Fermi level. The above observation and the presence of two values of $V$ for which $\mathcal{N}_R =0$ suggests that the transfer efficiency goes to zero when the resonant levels of the microlattice are aligned with $E_F$. This behavior shares similarities with the Thouless pump where the charge pumped through a non-interacting resonant level (for instance the energy level of a quantum dot) is zero on resonance \cite{fazio2005}. The above arguments show that gate potentials provide important control parameters to modulate the transfer efficiency of the peristaltic pump. The control parameter $V$ belongs to a relevant operation interval (defining the domain of Fig. \ref{fig:fig5}) limited by two extreme conditions. From one side the small-$V$ limit (i.e. $V\ll E_F$) cannot be considered since we are interested in describing two weakly coupled atomic reservoirs, while, on the other hand, the large-$V$ limit determines the complete decoupling of the two atomic reservoirs and consequently leads to a vanishing value of the pump efficiency.\\
The transfer efficiency of the pump can also be altered by changing the barrier width $2W$, without changing the microlattice spacing controlled by $Ka$. This analysis has been performed in Figure \ref{fig:fig6}, where the $\mathcal{N}_R$ \textit{vs} $W$ curves are shown.
\begin{figure}[h]
\includegraphics[clip,scale=0.8]{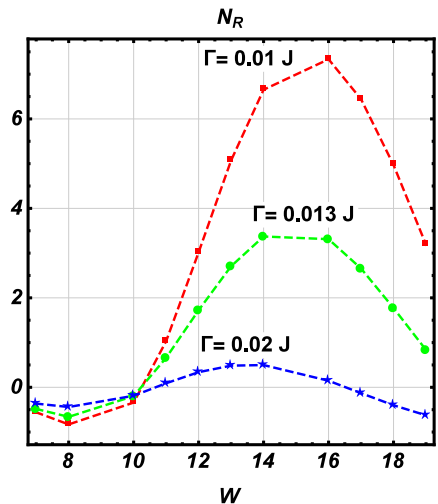}
\caption{Number of atoms $\mathcal{N}_R$ pumped in a pumping cycle as a function of the barrier half width $W$ (expressed in number of lattice sites) obtained by fixing the model parameters as follows: $\epsilon=2J$, $\hbar \omega/J=5\cdot 10^{-3}$, $V=0.37J$, $U_P=0.14J$, $Ka=0.8$, $n_0=40$, $N=80$, $E_F \approx 0.382 J$, $k_B T=0.1 \cdot E_F$ and $\Gamma=0.01J$ (box, $\square$), $\Gamma=0.013J$ (circle, $\bigcirc$), $\Gamma=0.02J$ (star, $\star$).}
\label{fig:fig6}
\end{figure}
The $\mathcal{N}_R$ \textit{vs} $W$ curves clearly show that the efficiency of the pump is suppressed when the mean free path $\ell$ (which is related to $\Gamma$) is lowered. In particular, setting the barrier width $2W=32$, we find $\mathcal{N}_R \approx 7.5$ with $\ell/(2W) \approx 1.90$ ($\Gamma=0.01J$), $\mathcal{N}_R \approx 3.2$ with $\ell/(2W) \approx 1.46$ ($\Gamma=0.013J$), while $\mathcal{N}_R \approx 0$ when $\ell/(2W) \approx 0.95$ ($\Gamma=0.02J$). These results put in evidence the quantum nature of the peristaltic mechanism and its fragility against the atoms finite mean free path. Indeed, when the particles mean free path $\ell$ becomes comparable with or shorter than the barrier width $2W$ the transfer efficiency of the pump is subject to the detrimental effects induced by decoherence. The latter behavior is known to occur in quantum machines or quantum motors. All the curves in Fig. \ref{fig:fig6} show that the transfer efficiency, measured by $\mathcal{N}_R$, is a growing function of $W$ for $W \lesssim 14$, while a decreasing behavior is observed when $W \gtrsim 14$. Two mechanisms are simultaneously at work and determine the observed behavior. On one side, changing the barrier width $2W$ changes the distance of the microlattice resonant levels from the Fermi energy. Since the microlattice spectrum is altered by $W$, the system can be driven on resonance by modifying the barrier width, the latter condition implying a vanishing pumping efficiency.
\begin{figure}[t]
\includegraphics[clip,scale=0.8]{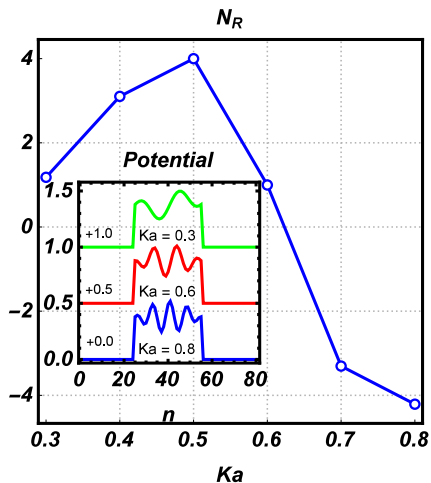}
\caption{Main panel: Number of atoms $\mathcal{N}_R$ pumped in a pumping cycle as a function of $Ka$ obtained by fixing the model parameters as follows: $\epsilon=2J$, $\hbar \omega/J=5\cdot 10^{-3}$, $V=0.385J$, $U_P=0.14J$, $W=15$, $n_0=40$, $N=80$, $E_F \approx 0.382 J$, $k_B T=0.1 \cdot E_F$ and $\Gamma=0.01J$. Inset: Potential profiles at the initial time $t=0$ obtained for $Ka=0.8$ (lower curve), $Ka=0.6$ (middle curve), $Ka=0.3$ (upper curve). A positive offset (specified close to the curves) has been added to the potential profiles.}
\label{fig:fig7}
\end{figure}
This mechanism induces an oscillating behavior of $\mathcal{N}_R$ with the microlattice length $2W$.
A second mechanism is related to the fact that an increasing barrier width produces a microlattice potential with an increased number of minima. As a consequence the microlattice region behaves like a multimodes one-dimensional channel characterized by an increased local density of states. Under these conditions the transfer efficiency of the pump is amplified. On the other hand, too long channels ($W>14$) are prone to decoherence effects and present a poor transfer efficiency.\\
An alternative way to change the number of minima of the microlattice potential consists in modifying the parameter $Ka$, while keeping fixed the barrier width. According to the same mechanism discussed before, the microlattice spectrum parametrically depends on $Ka$ and thus the resonance condition can be altered changing this parameter. Thus an oscillating behavior is expected for the $\mathcal{N}_R$ \textit{vs} $Ka$ curves. This expectation is confirmed by the results in Fig. \ref{fig:fig7}, where the $\mathcal{N}_R$ \textit{vs} $Ka$ curve has been reported. When $Ka$ is varied in the range $[0.3,0.8]$ the number of microlattice minima changes from $2$ to $4$ (see the inset of Fig. \ref{fig:fig7}). However, differently from the case presented in Fig. \ref{fig:fig6}, the microlattice spacing is here altered by changing $Ka$. In particular, increasing $Ka$ reduces the microlattice spacing with the twofold effect of (i) increasing the effective tunnel coupling between states localized within adjacent minima; (ii) increasing the energy associated to states localized within a single microlattice minimum. The simultaneous action of these spectrum deformations (parametrized by $Ka$) drives the system on resonance for $Ka \approx 0.62$, where the pump efficiency goes to zero.\\
\begin{figure}[t]
\includegraphics[clip,scale=0.8]{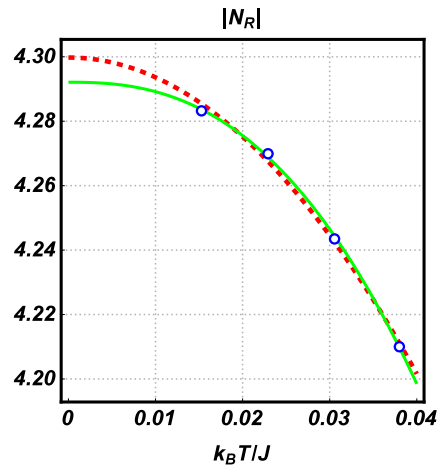}
\caption{Number of atoms $|\mathcal{N}_R|$ pumped in a pumping cycle as a function of the system temperature $k_B T/J$ obtained by fixing the model parameters as follows: $\epsilon=2J$, $\hbar \omega/J=5\cdot 10^{-3}$, $V=0.385J$, $U_P=0.14J$, $Ka=0.8$, $W=15$, $n_0=40$, $N=80$, $E_F \approx 0.382 J$ and $\Gamma=0.01J$ (empty circles). The full line represents the model $|\mathcal{N}_R(T)|=|\mathcal{N}_R(0)|(1-\kappa (k_B T/J)^{n})$ with the best fit parameters $|\mathcal{N}_R(0)|\approx 4.29$, $\kappa \approx 68.2$, $n=2.5$, while the dashed line is obtained considering $|\mathcal{N}_R(T)|=|\mathcal{N}_R(0)|(1-\kappa (k_B T/J)^{2})$ and $|\mathcal{N}_R(0)|\approx 4.3$, $\kappa \approx 14.3$.}
\label{fig:fig8}
\end{figure}
The pump efficiency is also affected by the system temperature which introduces detrimental effects related to a thermal smearing phenomenon. The efficiency of the system as a function of the temperature $T$ has been analyzed in Fig. \ref{fig:fig8}, where different phenomenological models of thermal-induced efficiency loss are extracted from the outcomes of the numerical simulations (empty circles). The analysis suggests that the thermal-induced efficiency loss follows the low-temperature relation $(|\mathcal{N}_R(0)|-|\mathcal{N}_R(T)|)/|\mathcal{N}_R(0)| \propto T^{n}$ with $n \sim 2-2.5$. This behavior is analogous to the one reported in Ref. \cite{fioretto2008} in the context of a finite-temperature quantum pumping theory. The above arguments show that the actual experimental temperatures within the typical range $T/T_F \lesssim 0.1$ do not provide severe limitations to the transfer efficiency of the pump like those induced by finite life time effects.

\section{Conclusions}
\label{sec:Conclusions}
We have proposed the concept of a \textit{peristaltic quantum pump} which is a quantum machine whose working principle is alternative with respect to the celebrated Thouless pump. This class of quantum devices generates a particles flux as the effect of the sliding motion of a finite-size microlattice placed in a one dimensional conduction channel. We have formulated a one-dimensional tight-binding Hamiltonian model for this quantum machine and the outcomes of the model have been analyzed within a lattice Green's function formalism on the Keldysh contour. The mathematical treatment allows the analysis of arbitrary peristaltic potentials and is particularly appealing to describe current experiments with fermionic cold atoms (e.g. $^{6}\mathrm{Li}$ atoms) where the mesoscopic microlattice can be obtained using light at $532$ nm holographically shaped by a digital micromirror device, as proposed in Ref. \cite{lebrat2017}. The pump observables, like e.g. the pumped atoms per cycle (efficiency), have been studied as a function of the pump parameters such as the pumping frequency, the width of the pumping potential, the atoms mean free path and the system temperature. Space-resolved quantities such as the bond currents generated by the peristaltic motion of the pump along the channel have also been derived. We have performed numerical simulations of the active region of the system, i.e. the region where the current is generated, and we have proven that the bond current is induced in the vicinity of the peristaltic potential and decays within few Fermi wavelengths inside the reservoirs. The analysis of the transfer efficiency (i.e. $\mathcal{N}_R$) as a function of the static potential $V$, of the width of the microlattice $2W$, of the phase profile $Ka$ evidences an oscillating behavior characterized by changes of sign. We have shown that the mentioned behavior, similarly to that of a Thouless pump, originates from the deformation of the microlattice spectrum caused by the variation of the peristaltic potential parameters. Changes of sign of the efficiency are observed around special points of the parameters space for which one resonant level of the microlattice spectrum is aligned with the system Fermi level. These characteristics can be used to maximize or invert the particle flux around a given working point and can have a practical relevance. We have characterized the loss of efficiency of the pump induced by temperature and inelastic scattering events. While the actual experimental temperature does not provide relevant limitations to the pump efficiency, inelastic events dramatically change the transfer efficiency of the system. We have quantified the inelastic scattering rate through the particle mean free path $\ell$. Severe limitations to the efficiency have been obtained for microlattice length $2W$ comparable to $\ell$. The detrimental role of decoherence evidences the fragility of a quantum machine against phase breaking events and is an important figure of merit for the experiment.

\section*{Acknowledgments}
The authors acknowledge discussions with J.-Ph. Brantut, L. Corman, T. Esslinger, T. Giamarchi, and D. Husmann.

\appendix
\section{The bare Green's functions and unperturbed density of states}
\label{app:gfDos}
In the absence of single-particle potentials the Hamiltonian of the problem coincides with $H_0$, which defines a time-independent problem. The retarded and advanced Green's functions admit the following definitions:
\begin{eqnarray}
&&g^{R}_{nm}(t,t')=-i\theta(t-t')\langle \{c_n(t),c^{\dagger}_m(t')\}\rangle \nonumber\\
&&g^{A}_{nm}(t,t')=i\theta(t'-t)\langle \{c_n(t),c^{\dagger}_m(t')\}\rangle.
\end{eqnarray}
Hereafter we derive the explicit expression of the retarded Green's function. Using the Heisenberg equation of motion $i\hbar \partial_{t} \mathcal{O}=[\mathcal{O},H_0]$ for the generic operator $\mathcal{O}$ we get
the following Fourier-transformed equation of motion for $g^{R}_{nm}(E)$:
\begin{eqnarray}
\label{eq:EOM-retgf0}
(E-\epsilon+i0^{+})g^{R}_{nm}(E)=\delta_{nm}-J \sum_{r=\pm 1}g^{R}_{n+r,m}(E),
\end{eqnarray}
where invariance under time-translation has been exploited. Due to the translational invariance with respect to the space, using the normalized eigenfunctions of the translation operator $\phi_{k}(n)=e^{ikan}/\sqrt{N}$, obeying the completeness relation $\sum_{k}\phi^{\ast}_{k}(n)\phi_{k}(m)=\delta_{nm}$, one can write the solution of Eq. (\ref{eq:EOM-retgf0}) according to the expansion:
\begin{eqnarray}
g^{R}_{nm}(E)=\frac{1}{N}\sum_{k}e^{i ka (n-m)}g^{R}(k,E),
\end{eqnarray}
with $g^{R}(k,E)=1/(E-E_k+i0^{+})$, while $E_{k}=\epsilon-2J \cos(ka)$ represents the dispersion relation of fermions delocalized over a one dimensional lattice. For a finite system, with $N$ lattice sites, the particle momentum takes the discrete values $ka=\frac{2\pi n}{N}$ with $|n|\leq (N-1)/2$. In the $N\rightarrow\infty$ limit, a continuum of energy levels is formed and the retarder Green's function can be approximated by the integral over the Brillouin zone:
\begin{eqnarray}
g^{R}_{nm}(E)=\frac{1}{2\pi}\int_{-\pi}^{\pi}d\theta \frac{e^{i\theta (n-m)}}{E-\epsilon+2J\cos(\theta)+i0^{+}},
\end{eqnarray}
which can be exactly solved using the residues method. Accordingly, we obtain:
\begin{eqnarray}
g^{R}_{nm}(E)=\frac{i}{2J}\frac{\Big(z-i\sqrt{1-z^2}\Big)^{|n-m|}}{\sqrt{1-z^2}}
\end{eqnarray}
with $z=(E+i0^{+}-\epsilon)/(2J)$, $J>0$. A finite life-time can be phenomenologically included making the substitution $i0^{+}\rightarrow i\Gamma$. Moreover the local density of states, $LDOS(E,n)=-Im[g^{R}_{nn}(E)]/\pi$, can be easily computed with the result:
\begin{eqnarray}
LDOS(E,n)=\frac{1}{2\pi J}\frac{1}{\sqrt{1-(\frac{E-\epsilon}{2J})^2}}.
\end{eqnarray}
For the sake of completeness, we also observe that the advanced Green's function is obtained by the retarded part according to the relation $g^{A}_{nm}(E)=[g^{R}_{mn}(E)]^{\ast}$, while in presence of time-dependent potentials we have $g^{A}_{nm}(E_1,E_2)=[g^{R}_{mn}(E_2,E_1)]^{\ast}$.

\section{Lesser Green's function of the unperturbed problem $\hat{g}^{<}(t,t')$}
\label{app:lesser}
The lesser Green's function in the absence of potential ($\hat{V}(t)=0$) can be computed using the definition given in Equation (\ref{eq:lesser}) with the following eigenfields expansion:
\begin{eqnarray}
c_n(t)=\sum_{k}\frac{1}{\sqrt{N}}e^{ikan}c_k e^{-iE_k t/\hbar},
\end{eqnarray}
where $E_{k}=\epsilon-2J \cos(ka)$ represents the energy eigenvalues of the unperturbed eigenfields $c_k$. After direct computation, using the relation $\langle c_{k}^{\dagger}c_{k'}\rangle=\delta_{kk'}f(E_k)$, we obtain:
\begin{eqnarray}
g_{nm}^{<}(t,t')=\frac{i}{N}\sum_{k}f(E_k)e^{-iE_k(t-t')/\hbar}e^{ika(n-m)}.
\end{eqnarray}
The particle density of the unperturbed system is simply given by $-ig^{<}_{nn}(t,t)=\sum_{k}f(E_k)/N$, with $f(E_k)$ the Fermi-Dirac distribution. Furthermore, the Fourier transform of $g_{nm}^{<}(t,t')$ is given by the expression:
\begin{eqnarray}
&&g_{nm}^{<}(E_1,E_2)=\\\nonumber
&=&\frac{(2\pi \hbar)^2 i}{N}\sum_{k}f(E_k)e^{ika(n-m)}\delta(E_1-E_k)\delta(E_1-E_2).
\end{eqnarray}

\section{Langreth theorem of analytic continuation and Keldysh equations}
\label{app:langreth}
Let us project on the real-time axis the contour correlation:
\begin{eqnarray}
\hat{C}(\tau,\tau')=\int_{\mathcal{C}_{K}}d\tau_a \ \hat{A}(\tau,\tau_a)\hat{B}(\tau_a,\tau').
\end{eqnarray}
According to our projection rule the above expression corresponds to the matrix equation:
\begin{widetext}
\begin{eqnarray}
                             &&  \left(
                               \begin{array}{cc}
                                 \hat{C}^{R}(t,t') &\hat{C}^{<}(t,t') \\
                                 0 & \hat{C}^{A}(t,t') \\
                               \end{array}
                             \right)=\int dt_a  \left(
                               \begin{array}{cc}
                                 \hat{A}^{R}(t,t_a) & \hat{A}^{<}(t,t_a) \\
                                 0 & \hat{A}^{A}(t,t_a) \\
                               \end{array}
                             \right) \left(
                               \begin{array}{cc}
                                 \hat{B}^{R}(t_a,t') & \hat{B}^{<}(t_a,t') \\
                                 0 & \hat{B}^{A}(t_a,t') \\
                               \end{array}
                             \right)=\\\nonumber
                             &=&\left(
                               \begin{array}{cc}
                               \int dt_a  \hat{A}^{R}(t,t_a) \hat{B}^{R}(t_a,t') & \int dt_a [\hat{A}^{R}(t,t_a)\hat{B}^{<}(t_a,t')+\hat{A}^{<}(t,t_a)\hat{B}^{A}(t_a,t')] \\
                                 0 & \int dt_a \hat{A}^{A}(t,t_a)\hat{B}^{A}(t_a,t') \\
                               \end{array}
                             \right),
\end{eqnarray}
\end{widetext}
whose components immediately give the following Langreth rules:
\begin{eqnarray}
\label{eq:langreth}
\hat{C}^{R/A}(t,t')&=&\int dt_a  \hat{A}^{R/A}(t,t_a) \hat{B}^{R/A}(t_a,t')\\
\hat{C}^{<}(t,t')&=& \int dt_a \Big[\hat{A}^{R}(t,t_a)\hat{B}^{<}(t_a,t')+\nonumber\\
&+&\hat{A}^{<}(t,t_a)\hat{B}^{A}(t_a,t')\Big]\nonumber.
\end{eqnarray}
The Lengreth relations given in Eq. (\ref{eq:langreth}) can be expressed using the shortened notation:
\begin{eqnarray}
&&(AB)^{R/A}=A^{R/A}B^{R/A}\\\nonumber
&&(AB)^{<}=A^{R}B^{<}+A^{<}B^{A},
\end{eqnarray}
which evidences the matrix structure in the Keldysh space. Langreth rules can be generalized to the product $ABC$ of three correlation functions:
\begin{eqnarray}
(ABC)^{R/A}&=&A^{R/A}B^{R/A}C^{R/A}\\\nonumber
(ABC)^{<}&=&(AB)^{R}C^{<}+(AB)^{<}C^{A}=\\\nonumber
&=&A^{R}B^{R}C^{<}+A^{R}B^{<}C^{A}+A^{<}B^{A}C^{A}.
\end{eqnarray}
When Langreth theorem is applied to the contour Dyson equation given in (\ref{eq:dysonK}), we obtain:
\begin{eqnarray}
G^{R/A}&=&g^{R/A}+g^{R/A}VG^{R/A}+g^{R/A}\Sigma^{R/A} G^{R/A}\nonumber\\
G^{<}&=&g^{<}+g^{R}VG^{<}+g^{<}VG^{A}+\\\nonumber
&+&g^{R}\Sigma^{R} G^{<}+g^{R}\Sigma^{<} G^{A}+g^{<}\Sigma^{A} G^{A},
\end{eqnarray}
where the fact that the one-body potential is diagonal in Keldysh space ($V^{<}=0$ and $V^{R/A}=V$) has been used in the derivation.
The infinite order iterate of the above equations leads to the following explicit equation for $G^{<}$:
\begin{eqnarray}
G^{<}&=&[1+G^{R}(V+\Sigma^{R})]g^{<}[1+(V+\Sigma^{A})G^{A}]+\nonumber\\
&+&G^{R}\Sigma^{<} G^{A},
\end{eqnarray}
which is known as Keldysh equation.

\section{Two-times Fourier transform}
\label{app:fourier}
Along this work the two-times Fourier transform of a generic correlation function $G$ is defined by the following relations:
\begin{eqnarray}
&&G(t_1,t_2)=\int\frac{dE_1 dE_2}{(2\pi \hbar)^2} G(E_1,E_2) e^{-\frac{i(E_1 t_1-E_2 t_2)}{\hbar}}\\
&&G(E_1,E_2)=\int dt_1 dt_2 G(t_1,t_2) e^{\frac{i(E_1 t_1-E_2 t_2)}{\hbar}}.
\end{eqnarray}
When the correlation only depends on $t_1-t_2$, i.e. $G(t_1,t_2)=\mathcal{F}(t_1-t_2)$, then
\begin{eqnarray}
G(E_1,E_2)=2\pi \hbar \delta(E_1-E_2)\int d\tau \mathcal{F}(\tau) e^{i\frac{E_1}{\hbar} \tau}.
\end{eqnarray}
This result is easily proved using the change of variables $\tau=t_1-t_2$ and $t=t_1+t_2$ inside the Fourier transform definitions.

\section{Transmission probability $T(E)$ through the microlattice potential}
\label{app:scattering microlattice}
\begin{figure*}
\includegraphics[clip,scale=0.75]{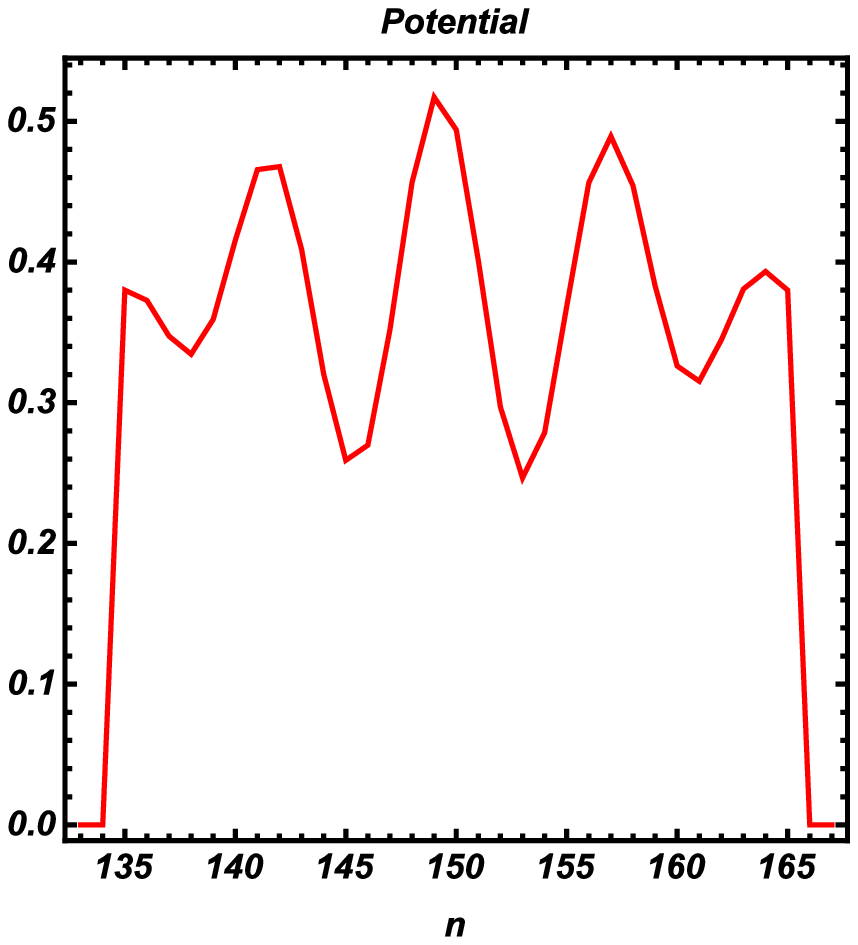}
\includegraphics[clip,scale=0.75]{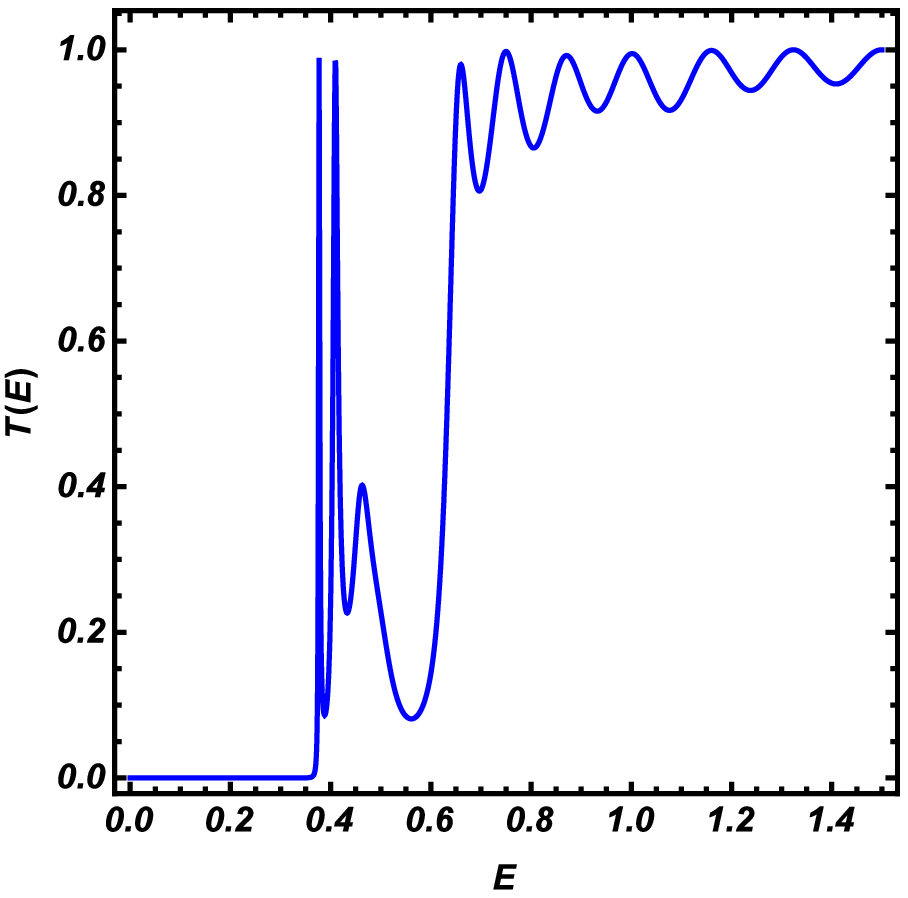}
\caption{Left Panel: Microlattice potential $\mathcal{U}_n$ defined by Eq. (23). Right Panel: Transmission probability $T(E)=|\tau(E)|^2$ through the potential $\mathcal{U}_n$ as a function of the energy $E$ of the scattering states. The energy is measured in units of the hopping $J$. Quantum states describing particles trapped inside the microlattice define a sort of sub-band which enables resonant transmission phenomena for $E \in[0.38J,0.5J]$. For $E>0.65J$ the potential barrier is overcome by the scattering particles and the transmission probability reaches its maximum value. Total reflection is observed for $E<0.38J$.}
\label{fig:figtransmission}
\end{figure*}

This analysis shows the relevance of quantum states describing particles trapped inside the minima of the microlattice potential. We study a scattering problem through the static potential given in Eq. (23) and reproduced in Fig. \ref{fig:figtransmission} (left panel). The asymptotic scattering states $\psi_{L/R}(n)$ belonging to the (translational-invariant) left and right atomic cloud can be written as:
\begin{eqnarray}
\psi_L(n)&=&\exp(i k_E a n)+r(E)\exp(-i k_E a n)\\\nonumber
\psi_R(n)&=&\tau(E)\exp(i k_E a n),
\end{eqnarray}
with the lattice index $n \notin [135,165] \equiv \mathcal{A}$. Assuming dispersion relation $E=2J(1-\cos(ka))$ for scattering states belonging to the left/right atomic cloud, the particle momentum is given by $k_E a=\arccos(\frac{2J-E}{2J})$, while $E \in[0,4J]$. The transmission properties of the system are given by $\tau(E)$, while the particles flux conservation implies $|\tau(E)|^2+|r(E)|^2=1$. Transmission and reflection coefficients, namely $\tau(E)$ and $r(E)$, are obtained according to the following procedure.
Let us call $\mathcal{U}_n$ the static potential given in Eq. (23). Within this notation, fixing the energy $E$ of the scattering states, the tight-binding (time-independent) Schr\"{o}dinger equation for the lattice wavefunction $\Psi_n$ implies:
\begin{eqnarray}
\label{eq:tbschro}
(2J+\mathcal{U}_n-E)\Psi_n-J(\Psi_{n-1}+\Psi_{n+1})=0.
\end{eqnarray}
By varying the lattice index $n$ in the interval $[134,166]$ and setting the following boundary conditions:
\begin{eqnarray}
\Psi_{133}&=& \psi_{L}(133)\\\nonumber
\Psi_{134}&=& \psi_{L}(134)\\\nonumber
\Psi_{166}&=& \psi_{R}(166)\\\nonumber
\Psi_{167}&=& \psi_{R}(167),
\end{eqnarray}
Eq. (\ref{eq:tbschro}) generates a set of $33$ equations with respect to the unknowns $\{\tau(E),r(E), \Psi_{135},...,\Psi_{165}\}$. The numerical solution of the problem allows us to obtain the transmission probability $T(E)=|\tau(E)|^2$, which is shown in Fig. \ref{fig:figtransmission} (right panel). The analysis of the transmission probability reveals the presence of four distinct scattering regimes: (a) Total reflection, occurring for scattering states with energy $E<0.38J$; (b) Resonant transmission, taking place for $E \in [0.38J,0.5J]$; (c) Over the barrier reflection for $E \approx 0.55J$; (d) Over the barrier full transmission for $E>0.65J$. Interestingly, the resonant transmission region in (b) originates from the merging of three resonant peaks with different width, which are the fingerprint of quantum states describing particles trapped inside the microlattice minima. The peculiar behavior of the transmission probability mimics the presence of a small sub-band associated to the microlattice potential. The arguments mentioned above show the relevance of the resonant states (discussed in (b)) in determining the transport properties of the system especially when the system chemical potential is close to $E_F \approx 0.38J$.

\end{document}